\documentclass[superscriptaddress,twocolumn,aps,pra,10pt,longbibliography]{revtex4-1} 
\usepackage{graphicx}
\usepackage{epsfig}
\usepackage[caption=false]{subfig}
\usepackage{verbatim}
\usepackage{amssymb}
\usepackage{amsmath}
\usepackage{amsfonts}
\usepackage{refcount}
\usepackage{bm}
\usepackage{hyperref}
\usepackage{color}
\usepackage[margin=.7in,nofoot]{geometry}
\usepackage{hyperref}
\hypersetup{
    colorlinks=true,       
    linkcolor=cyan,          
    citecolor=magenta,        
    filecolor=magenta,      
    urlcolor=cyan,           
    runcolor=cyan
}
\usepackage{algorithm}
\usepackage{algpseudocode}

\newcommand{\bea}{\begin{eqnarray}}
\newcommand{\eea}{\end{eqnarray}}
\newcommand{\beq}{\begin{equation}}
\newcommand{\eeq}{\end{equation}}
\newcommand{\norm}[1]{\left\lVert#1\right\rVert}

\newcommand{\nsol}{D}

\newcommand{\appdwave}{ the Supplementary Information}
\newcommand{\appcsa}{ the Supplementary Information}

\begin{document}

\title{Advantages of Unfair Quantum Ground-State Sampling}

\author{Brian Hu Zhang}
\affiliation{Stanford University, Stanford, California 94305, USA}

\author{Gene Wagenbreth}
\affiliation{Cray, Seattle, WA 98164, USA}

\author{Victor Martin-Mayor}
\affiliation{Departamento de F\'isica Te\'orica I, Universidad Complutense, 28040 Madrid, Spain}
\affiliation{Instituto de Biocomputaci\'on y F\'isica de Sistemas Complejos (BIFI), Zaragoza, Spain}

\author{Itay Hen}
\email{itayhen@isi.edu}
\affiliation{Information Sciences Institute, University of Southern California, Marina del Rey, California 90292, USA}
\affiliation{Department of Physics and Astronomy and Center for Quantum Information Science \& Technology, University of Southern California, Los Angeles, California 90089, USA}

\date{\today}

\begin{abstract}
The debate around the potential superiority of quantum annealers over their classical counterparts has been ongoing since the inception of the field. Recent technological breakthroughs, which have led to the manufacture of experimental prototypes of quantum annealing optimizers with sizes approaching the practical regime, have reignited this discussion. However, the demonstration of quantum annealing speedups remains to this day an elusive albeit coveted goal. We examine the power of quantum annealers to provide a different type of quantum enhancement of practical relevance, namely, their ability to serve as useful samplers from the ground-state manifolds of combinatorial optimization problems. We study, both numerically by simulating stoquastic and non-stoquastic quantum annealing processes, and experimentally, using a prototypical quantum annealing processor, the ability of quantum annealers to sample the ground-states of spin glasses differently than thermal samplers. We demonstrate that i) quantum annealers sample the ground-state manifolds of spin glasses very differently than thermal optimizers, ii) the nature of the quantum fluctuations driving the annealing process has a decisive effect on the final distribution, and  iii) the experimental quantum annealer samples ground-state manifolds significantly differently than thermal and ideal quantum annealers. We illustrate how quantum annealers may serve as powerful tools when complementing standard sampling algorithms.
\end{abstract}

\maketitle


\section{Introduction}
Many problems of practical importance may be cast as a task of finding all
the minimizing configurations, or ground-states, of a given cost
function.  Examples are numerous---among them are SAT
filtering\cite{Douglass2015}, hardware fault detection and the
verification and validation (V\&V) of safety-critical cyber-physical
systems\cite{Pudenz:2013kx}, to mention a few.  
In the V\&V of safety-critical cyber-physical systems for instance, one is
concerned with testing whether a given piece of software contains a
bug. This problem can naturally be cast as a constraint satisfaction
problem\cite{Pudenz:2013kx} (equivalently, as an optimization
problem of the Ising-type) where finding as many of the bugs as possible is critical to the
success of the mission. Similar scenarios occur in circuit fault
detection where each solution corresponds to a potential discrepancy in the
implementation of a circuit and where all discrepancies must be found.

The listing of all solutions of a given cost function
is, however, generally an intractable task for standard algorithms (it is a
problem in the complexity class \#P). This is not only because of the
difficulty involved in finding an
optimum\cite{papadimitriou2013combinatorial}, but also because of the
sheer number of ground-states of the problem which may grow exponentially with input size (a property known to physicists as a non-vanishing entropy density\cite{pauling:35}). Furthermore, the
energy landscapes of certain cost functions are known to bias
heuristic optimizers, as well as provable solvers, towards certain
solutions and away from others\cite{bastea:98}.  Thus, the practical importance of sampling from the
set of ground-states of intricate cost functions in qualitatively
diverse manners is immense---both from the theoretical point of view and for practical reasons. In the context of V\&V for instance, one hopes that employing a suite of qualitatively dissimilar sampling algorithms will unearth
nonidentical or even disjoint sets of solutions, leading eventually to the discovery of much larger sets of bugs. 

Recent technological breakthroughs that have made 
experimental programmable quantum annealing (QA) optimizers containing
thousands of quantum bits\cite{johnson:11,berkley:13} available, have rekindled
the interest in annealers as a revolutionary new approach to finding the minimizing assignments of discrete
combinatorial cost functions.  Quantum annealers\cite{kadowaki:98,farhi:01}
provide a unique approach to finding the ground-states of discrete
optimization problems, utilizing gradually decreasing quantum
fluctuations to traverse barriers in the energy landscape in
search of global optima, a mechanism commonly believed to have no
classical
counterpart\cite{Finnila1994343,Brooke30041999,kadowaki:98,farhi:01,santoro:02,RevModPhys.80.1061,PhysRevB.39.11828}.
In the context of ground-state sampling, quantum annealers thus offer the exciting possibility of discovering minimizing assignments that cannot be reached in practice with
standard algorithms, potentially offering unique advantages over traditional algorithms for solving problems of practical importance.

Here, we put this hypothesis to the test by directly addressing the question of whether quantum annealers can sample
the ground-state set of optimization problems differently than their
classical counterparts. We further examine the potential inherent in quantum annealers to serve as useful
tools in practical settings.  We demonstrate, both via numerical
simulations and experimentally, by testing a $512$-qubit D-Wave Two
quantum annealing optimizer\cite{johnson:11,berkley:13}, that
quantum annealers not only produce different distributions over the
set of ground-states than simulated thermal annealing optimizers, but
that they offer an additional dimension of tunability that does not
necessarily have a classical counterpart. 
Finally, we show that when used in conjunction with
existing standard ground-state-sampling or solution counting
algorithms, quantum annealers may offer certain unique advantages that may not be otherwise achievable.

\section{Sampling the ground-state manifold of spin glasses}


 

Similar to standard classical algorithms, quantum annealers---even ideal fully adiabatic ones held at zero temperature---when tasked with solving optimization problems will generally sample the solution space of optimization problems in a biased manner, producing certain ground-states more frequently than others. Unlike the bias exhibited by thermal algorithms, the uneven sampling of quantum annealers has its origins in the quantum nature of their dynamics: In standard quantum annealing protocols, one engineers a smoothly interpolating 
Hamiltonian between a simple `driver' Hamiltonian $H_d$ which provides the quantum fluctuations and a classical `problem'
Hamiltonian $H_p$ that is diagonal in the computational basis and whose ground-states encode the solutions of an optimization problem 
\begin{equation}\label{eq:hs}
H(s)=(1-s) H_d + s H_p \,,
\end{equation}
where $s(t)$ is a parameter varying smoothly with time from $0$ at
$t=0$ to $1$ at the end of the algorithm, at $t=\mathcal{T}$ [the type of problem and driver Hamiltonians we shall consider are given in Eqs.~\eqref{eq:Hp},~\eqref{eq:hdtf}
  and~\eqref{eq:def-hdnsq} in the next section].

\begin{figure}
{\includegraphics[width=0.9\columnwidth]{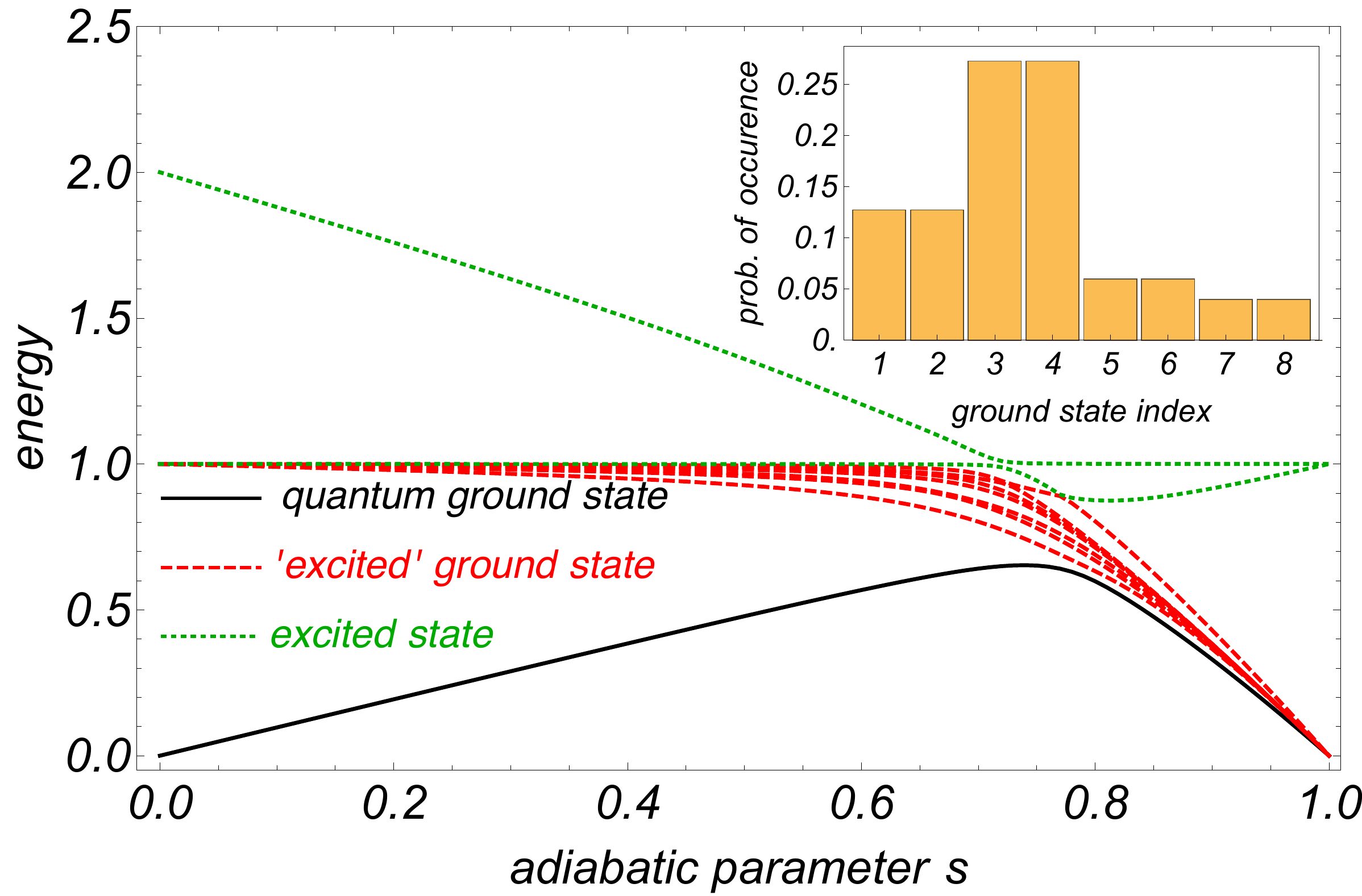}}
\caption{\textbf{Ten lowest energy levels of an $8$-qubit Hamiltonian interpolating between a transverse field driver Hamiltonian and a randomly generated Ising Hamiltonian.} The solid black line indicates the energy of the instantaneous ground-state. The dashed red and dotted green lines indicate excited states that lead to final ground-states and final excited states, respectively. {\bf Inset:} The probabilities for obtaining the various classical ground-states upon measuring the quantum ground-state at the end of the evolution in the computational basis.}
\label{fig:example}
\end{figure}

In the presence of degeneracy in the ground-state manifold of the
problem Hamiltonian (which corresponds to multiple minimizing assignments for the cost function), the adiabatic theorem ensures that the state
reached at the end of the adiabatic evolution in the limit $\lim_{s
  \to 1} H(s)$ is still uniquely defined. At the end of the QA
evolution, the final state corresponds to a specific
linear combination of the classical ground-states
\beq\label{eq:def-anneal-GS} |\psi_{GS}\rangle = \sum_{i=1}^{D} c_i |
\phi_i \rangle \eeq where $\{|\phi_1 \rangle,|\phi_2
\rangle,\ldots,|\phi_D \rangle\}$ is the set of $D$ classical ground
states, or minimizing configurations, of the optimization problem (see Fig.~\ref{fig:example} for an example). The
$\{|c_i|^2\}$ are the probabilities for obtaining each of these
classical ground-states upon computational-basis measurements at the
end of the anneal. These define a probability
distribution over the ground-state manifold and depend not only on
the structure of the problem Hamiltonian but also on the nature of the quantum
fluctuations provided by the driver Hamiltonian (from the point of
view of quantum perturbation theory, in the simplest case where first-order
perturbation theory breaks all degeneracies, the ground-state in Eq.~\eqref{eq:def-anneal-GS} is merely the ground-state of a
restricted driver Hamiltonian, specifically, the driver $H_d$ projected onto
the subspace spanned by the ground-states of $H_p$).
 
Since the distribution of minimizing configurations (henceforth,
the ground-state distribution, or GSD) generated by a quantum annealer is
\emph{intrinsically quantum}, a possibility arises that some quantum
GSDs cannot be efficiently generated by classical samplers. Moreover,
that the choice of driver Hamiltonian $H_d$ determines these GSDs, offers
a tunable handle, or an extra knob, that potentially produces a continuum of
probability distributions over the ground-state configurations. It is
therefore plausible to assume that certain classically suppressed
configurations, i.e., solutions that have very low probabilities of
being found via thermal or other classical processes, may have high
probabilities of being found or sampled with suitable choices of
driver Hamiltonians\cite{q-sig,q-sig2,matsuda,exponentiallyBiased}.  
In such cases, quantum annealers may be used to replace or \emph{complement} classical samplers, giving rise to a novel form of quantum enhancements.

To test whether quantum annealers indeed provide a
potentially powerful platform for achieving quantum enhancements for
the counting or listing of solutions of hard optimization problems, we study in detail their capabilities
to sample ground-state configurations differently than their classical
counterparts. As a testbed, we consider spin glasses: disordered,
frustrated spin systems\cite{young:98} that may be viewed as
prototypical classically hard (also called NP-hard) optimization
problems\cite{barahona:82} focusing, for reasons that will become
clear later, on problems whose ground-state configurations have been
computed in advance. These will be used to test the performance of
classical thermal annealers, comparing the outcomes of these against
the GSDs produced by ideal zero-temperature stoquastic as well as
non-stoquastic quantum annealers. We shall also compare the produced GSDs to those obtained by a prototypical
experimental quantum annealer---the D-Wave Two (DW2)
processor\cite{johnson:11,Bunyk:2014hb} which consists of an array of
superconducting flux qubits designed to solve Ising model instances
defined on the graph hardware via a gradually decreasing transverse
field Hamiltonian (further details, including a visualization of the
Chimera graph and the annealing schedule used to interpolate between
$H_d$ and $H_p$, are provided in \appdwave).
These comparisons will provide insight into the potential computational power inherent in quantum devices to assist traditional algorithm in finding all (or as many as possible) minimizing configurations of discrete optimization problems.

\section{Results: Thermal vs. quantum ground-state sampling}
\subsection{Setup\label{sec:setup}}
We generate random spin glass instances for which the enumeration of minimizing configurations is a feasible task. This a priori requirement will allow us to properly evaluate the sampling capabilities of the tested algorithms in an unbiased way. 
To do that, we consider optimization problems whose cost function is of the form:
\beq 
\label{eq:Hp} H_{p}=\sum_{\langle ij\rangle} J_{ij} s_i
s_j  \,.
\eeq 
The Ising spins, $s_i=\pm 1$, are the variables to be optimized over,
and the set of parameters $\{J_{ij}\}$ determines the cost function.  
To experimentally test the generated instances, we take $\langle ij\rangle$
to sum over the edges of an $N=504$-qubit Chimera graph---the hardware graph of the DW2 processor\cite{Bunyk:2014hb}. 

\begin{figure}[hbp]
{\includegraphics[width=0.99\columnwidth]{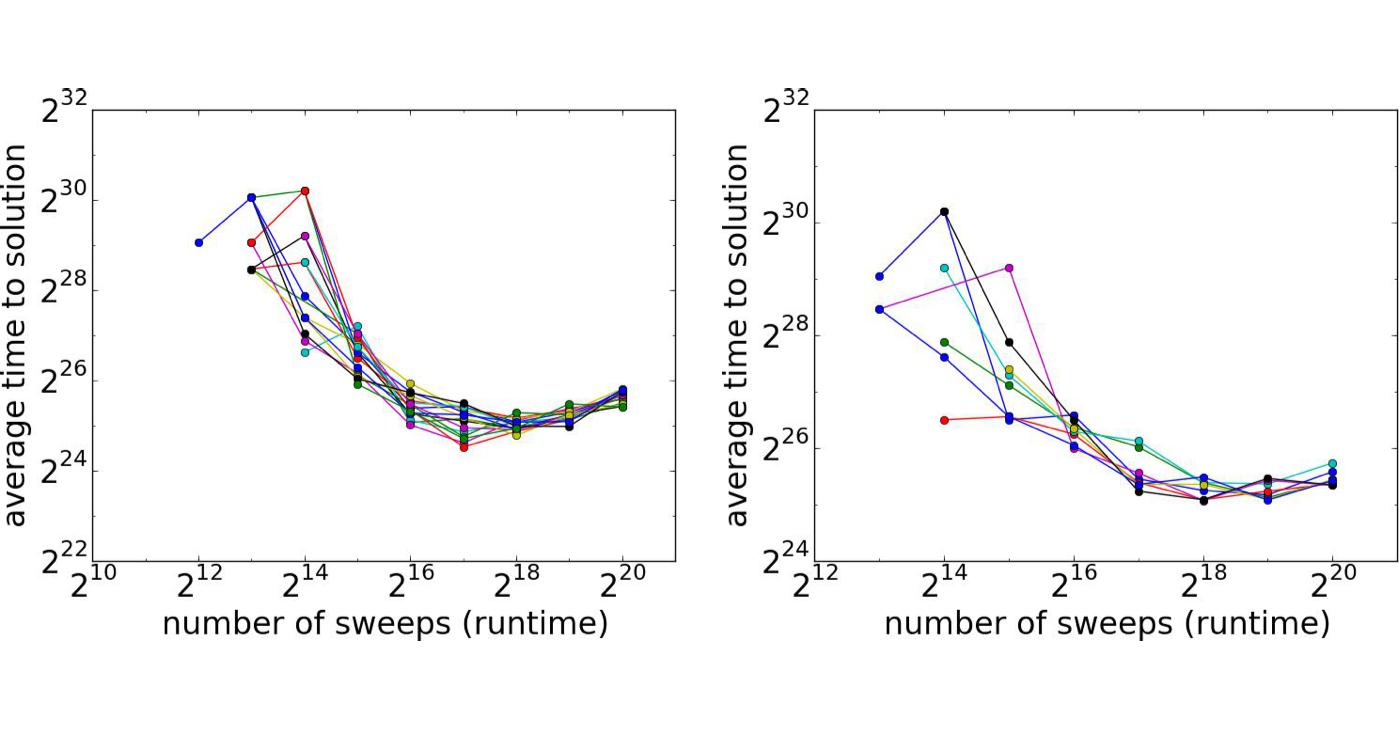}}
\caption{{\bf Simulated annealing average time to solution---the ratio of success probability to number of sweeps, vs. number of sweeps for different solutions (log-log scale)}. Each line represents a different solution. {\bf Left:} A $504$-bit instance with $16$ solutions. {\bf Right:} A $504$-bit instance with $4$ solutions.  For most instances, certain solutions are reached considerably sooner than others.}
\label{fig:sa_tts}
\end{figure}

The precise enumeration of all minimizing configurations of spin glass instances of more than a few dozen spins is generally an intractable task. We overcome this difficulty here by generating problem instances with \emph{planted solutions}\cite{hen:15,king:15}. As we discuss  in the {\bf{Methods}} section, the structure of planted-solution instances allows for the development of a constraint solving bucket algorithm capable of enumerating all minimizing assignments. 
By doing so, we obtain about $2000$ optimization $504$-bit spin glass instances, each with less than $500$ minimizing configurations, and for which we know {\emph{all}} ground-state configurations. This enables the accurate evaluation of the distributions of success probabilities of \emph{individual  minimizing  configurations}.
We compare the GSDs as obtained by several different algorithms:
\begin{enumerate}
\item \emph{Simulated annealing (SA)}---This is a well-known, powerful and generic heuristic solver\cite{kirkpatrick:83}.
Our SA algorithm uses single spin-flip Metropolis updates with a linear profile of inverse temperatures $\beta=T^{-1}$, going from $\beta_{\min}=0$ to $\beta_{\max}=20$ (with  $\beta$ updated after every Metropolis sweep over the lattice spins).
Figure~\ref{fig:sa_tts} provides the SA results for a couple of typical instances differing in number of minimizing configurations. The figure shows the average time for each individual minimizing bit assignment plotted as a function of number of sweeps on a log-log scale.
For any fixed number of sweeps, we find that the probability of
obtaining certain ground-state configurations may vary by orders of
magnitude. At first glance, the `unfair'
sampling of SA may seem to contradict the Boltzmann distribution 
which (in accordance with the assumption of equal a priori probability)
prescribes the same probability to same-energy configurations for thermalized systems. However, we note that when SA is used as an optimizer (as it is here), the number of sweeps is not large enough for thermalization to take place. As an optimizer, the number of SA Metropolis sweeps is chosen such that on average the time to find a ground-state is minimized. 
The minimum time-to-solution (for individual solutions) is  evident in Fig.~\ref{fig:sa_tts}, as is the convergence of all individual success probabilities into a single value in the limit of long annealing times, consistently with equipartition. 
\item\emph{Ideal zero-temperature quantum annealer with a transverse field driver (TFQA)}---We also consider an ideal, zero-temperature, fully adiabatic quantum annealer with a transverse-field driver Hamiltonian, namely
\beq\label{eq:hdtf}
H_d^{\text{TF}} =- \sum_i \sigma_i^x \,,
\eeq
where $\sigma_i^x$ is the Pauli X spin-$1/2$ matrix acting on spin $i$. 
The adiabatic process interpolates linearly between the above driver and the final Ising Hamiltonian (Eq.~[\ref{eq:Hp}]) as described in Eq.~(\ref{eq:hs}). The algorithm we devise to infer the quantum ground-state is discussed in detail in {\bf Methods}.
\item\emph{Ideal zero-temperature quantum annealer with a non-stoquastic driver (NSQA)}---As already discussed above, quantum annealers offer more control than thermal annealers over the generated fluctuations. This is because unlike thermal fluctuations, quantum fluctuations can be \emph{engineered} (see, e.g., Refs.\cite{dickson:12,crosson:15}). One thus expects different driver Hamiltonians to yield different probability distributions over the ground-state manifold.
Of particular interest are the so-called \emph{non-stoquastic} driver Hamiltonians, which cannot be efficiently simulated by classical algorithms. As a test case, we consider an additional quantum annealing process driven by:
\beq\label{eq:def-hdnsq}
H_d^{\text{NS}} =- \sum_i \sigma_i^x +\sum_{\langle i j \rangle} \tilde{J}_{ij} \sigma_i^x \sigma_j^x \,.
\eeq
To ensure that the driver in non-stoquastic, we choose the couplings $\tilde{J}_{ij}$ to be the same as the $\sigma^z_i \sigma^z_j$ couplings of the problem Hamiltonian, but with an arbiltrarily chosen sign. 
\item\emph{Experimental D-Wave Two processor (DW2)}---We also feed the generated instances to the putative DW2 quantum annealing optimizer.
This device is designed to solve optimization problems by evolving a known initial configuration---the ground-state of a transverse field towards the ground-state of the classical Ising-model Hamiltonian of Eq.~(\ref{eq:Hp}). Each problem instance was run on the annealer for a minimum of $10^4$ anneals with each anneal lasting $20$ to $40 \mu$s, 
for a total of more than $10^{7}$ anneals. 
Each anneal ends up with a measurement in the computational basis yielding either an excited state or a classical ground-state which is subsequently recorded and which is later used to construct a GSD. To overcome the inhomogeneity of the processor as well as other systematic errors, each anneal is carried out with a randomly generated gauge (see Ref.\cite{q108} for more details).
\end{enumerate}
\subsection{Distinguishing Probability Distributions}
We test whether the different algorithms that we consider produce significantly different GSDs, or probability distributions over the ground-state manifold, on the various spin glass instances. 
To distinguish between two distributions generated by two methods, at
least one of which is empirically estimated via experiment, we use a
bootstrapped version of the Kolmogorov-Smirnov (KS) test (see {\bf
  Methods}).

Table~\ref{table:properties} summarizes the results of the statistical tests, listing the fraction of instances with different GSDs between any two tested optimizers. As is evident from the table, the distributions generated by the various algorithms are in general significantly different from one another---
pointing to presumably different physical mechanisms generating them, namely thermal or quantum fluctuations of different sources. As we discuss later on, these pronounced differences in the GSDs can allow quantum annealers to serve as potentially powerful tools, when combined with standard techniques, for finding all (or as many as possible) minimizing configurations of combinatorial optimization problems.  

\begin{table}[hp]
\centering
\begin{tabular}{|c||c|c|c|c|}
\hline
  & SA & TFQA & NSQA & DW2 \\\hline \hline
SA & - &$67\%$ &$65\%$ & $93\%$ \\\hline
TFQA &  & - & 57\% & $99\%$ \\\hline
NSQA &  & &- & $99\%$ \\\hline
DW2 &  & & & - \\\hline
\end{tabular}
\caption{{\bf Fraction of instances with statistically significant differences in GSD between any two optimizers. Here, the $p$-value is set at $p=0.01$.}
}
\label{table:properties}
\end{table}

To quantify the utility of using a combination of two (or more)
methods for finding as many minimizing configurations as possible, we
define the `bias' $b({\bf p})$ of a GSD as 
\beq b({\bf
  p})=\frac{\nsol}{2(\nsol-1)}\sum_{i=1}^{\nsol} \left| p_i-1/\nsol \right|\,,
 \eeq 
where we denote probability distributions by
${\bf p}$, and where $p_i$ is the probability of obtaining ground-state $i$.
Here, a flat distribution corresponds to $b({\bf p})=0$, whereas an extremely
biased one for which all samples are multiples of the same configuration
yields $b({\bf p})=1$.
If $n$ applications of one optimization method yield a probability distribution ${\bf p}^{(1)}$ and $n$ applications of another yield ${\bf p}^{(2)}$, then a combined effort of $n/2$ samples from each will yield a GSD with a bias $b({\bf \bar{p}})$ where 
\hbox{${\bf \bar{p}} = \frac1{2}\left({\bf p}^{(1)}+{\bf p}^{(2)} \right)$}. 
We can therefore quantitatively measure the utility of a combination of two (or more) methods by comparing the bias $b({\bf \bar{p}})$ against that of any one method $b({\bf p}^{(1)})$ [or $b({\bf p}^{(2)})$]. The smaller the bias of the combination, the greater the utility of using the two methods in conjunction. 
Let us next examine the differences between specific pairs of methods in more detail . 
\subsection{SA vs. TFQA and NSQA}
\begin{figure}[htp]
{\includegraphics[width=0.99\columnwidth]{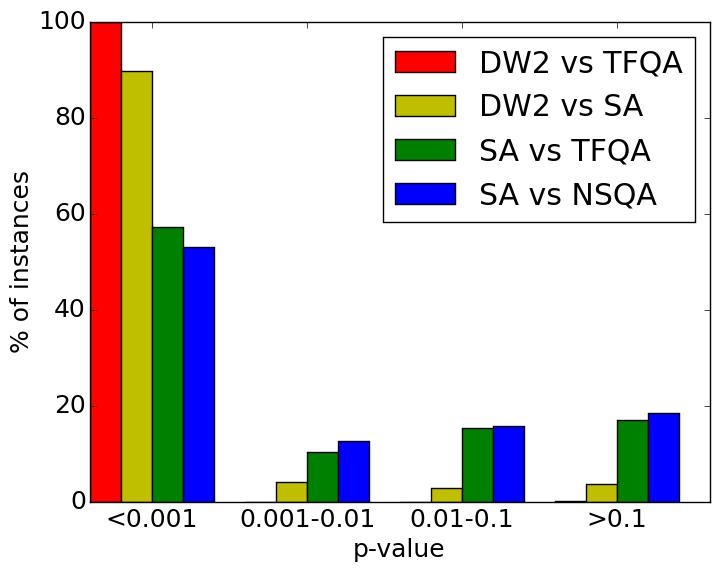}}
\caption{{\bf The $p$-values generated by the Kolmogorov-Smirnov tests to quantify the differences between pairs of algorithms}.}
\label{fig:pvalues}
\end{figure}

We first compare the GSDs obtained by thermal simulated annealing (SA) against those generated by the transverse-field and the non-stoquastic quantum annealers.
A bootstrapped KS test to decide whether they are significantly different suggests a difference that is significant at the $p < 0.01$ level in $67\%$ of the instances with TFQA and $65\%$ of the instances with NSQA. These results are summarized in Fig.~\ref{fig:pvalues}. 
In the vast majority of cases, there is a qualitative difference between the results produced by QA and those produced by SA. 

\begin{figure*}[htp]
{\includegraphics[width=0.99\textwidth]{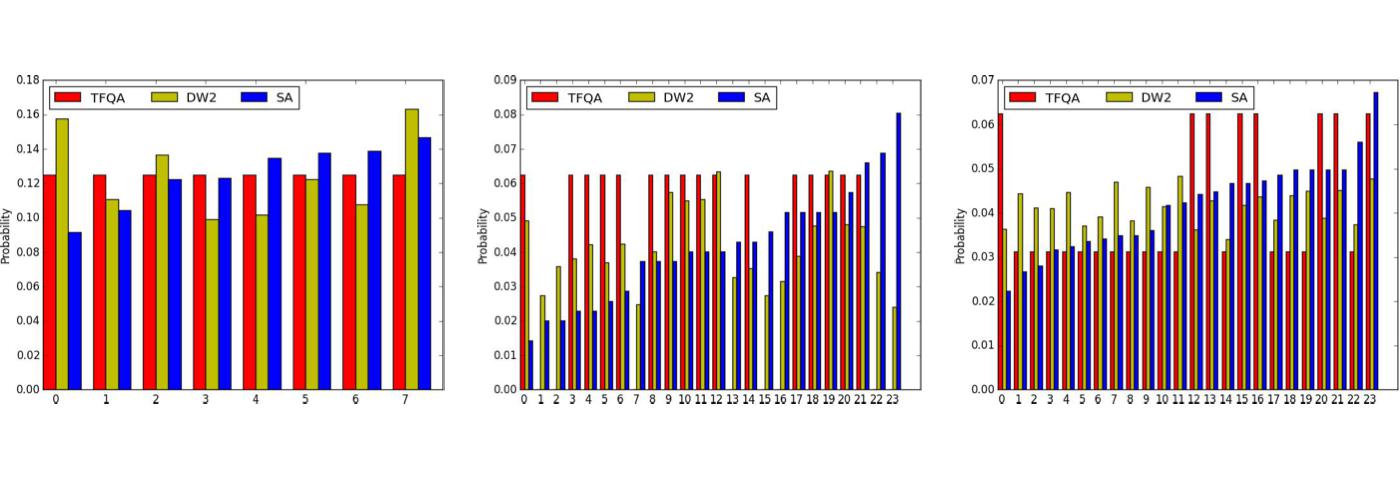}}
\caption{{\bf Three representative GSDs for simulated annealing (SA - blue), ideal transverse-field quantum annealer (TFQA - red) and the D-Wave Two experimental processor (DW2 - yellow).} In the left instance, probabilities for obtaining the various ground-states predict that all solutions are about equally likely. In the middle instance, we observe that those ground-states that our analytic TFQA predicts should appear more often, and do indeed appear more often in the experimental DW2 data. In the right instance, there is no clear relationship between the various algorithms.}
\label{fig:pdgd}
\end{figure*}

Figure~\ref{fig:pdgd} depicts the GSDs of several representative instances illustrating the little or no relationship between the thermal and TFQA or NSQA methods. In the left panel we find an instance for which both SA and TFQA produce similar GSDs. The middle and right instances show no clear relationship between SA and the other algorithms. 

\begin{figure}[htp]
{\includegraphics[width=0.99\columnwidth]{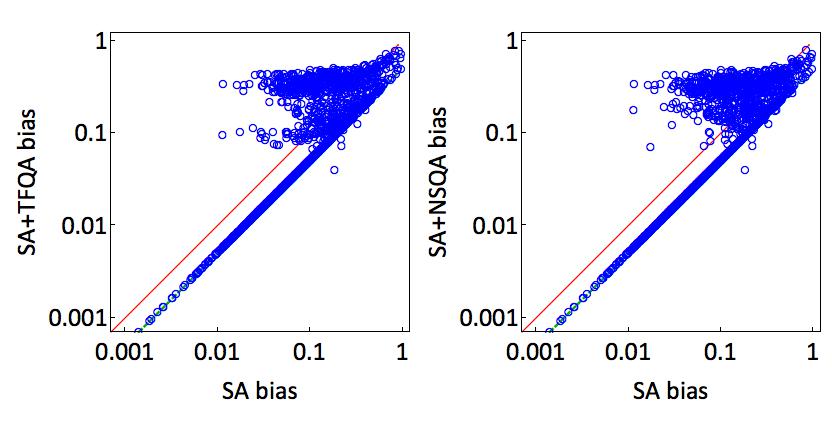}}\caption{
{\bf Scatter plots of biases of the tested instances' GSDs as obtained by SA vs. a combination of SA with ideal quantum annealers.} Left: SA vs. SA+TFQA. Right: SA vs. SA+NSQA. In the majority of cases, a combination of the methods leads a smaller overall bias, i.e., a lesser degree of unfairness. The solid red line is the `equal bias' $y=x$ line, whereas the dashed green $y=x/2$ line is the bias obtained when SA is combined with a flat, unbiased GSD, in which case the bias is halved.
}
\label{fig:SAbiasPlots}
\end{figure}

\begin{figure}[hbp]
{\includegraphics[width=0.99\columnwidth]{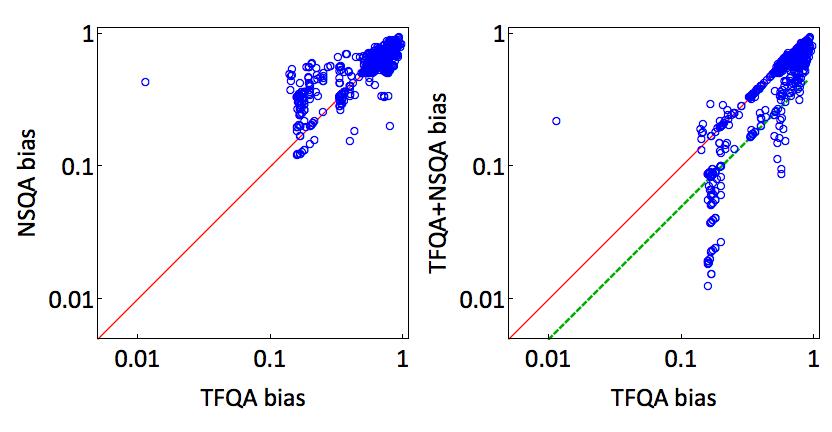}}
\caption{{\bf Scatter plots of biases of the tested instances' quantum GSDs.} Left: Biases of TFQA annealing vs. NSQA annealing showing that the non-stoquastic driver is slightly more biased in general than the transverse field driver (not shown in the figure are the completely unbiased instances which constitute about $57\%$ of the instances). Right: Biases of the TFQA GSDs vs. those of the combined application of TFQA and NSQA indicating that use of an additional non-stoquastic driver considerably reduces the bias of the GSDs.}
\label{fig:QAbiasPlots}
\end{figure}

We now ask whether using a quantum annealer \emph{together} with a simulated annealer has merit.
To that aim, we compare the SA bias $b_{\text{SA}}=b({\bf p}_{\text{SA}})$ against the biases of the combinations SA with TFQA and SA with NSQA. The results are summarized in Fig.~\ref{fig:SAbiasPlots}. In these scatter plots, any data point below the $y=x$ line indicates an advantage to using `assisted' ground-state sampling driven by quantum samplers. As can be clearly observed,  for most of the instances the bias of the combination is significantly lower than that of using SA alone. The median SA bias of $\langle b_{\text{SA}} \rangle =0.10(1)$ drops to $\langle b_{\text{SA+TFQA}} \rangle =0.075(0)$ and $\langle b_{\text{SA+NSQA}} \rangle =0.074(2)$ when assisted with TFQA and NSQA, respectively. Also noticeable is the $y=x/2$ line; data points on this line are obtained whenever SA is used in conjunction with a method that yields a flat distribution, in which case the bias is halved. We find, perhaps not surprisingly, that ideal zero-temperature annealing yields flat distributions for many of the tested instances [see also Fig.~\ref{fig:QAbiasPlots}(left)]. This is somewhat analogous to an ideal SA process where full thermalization is reached, in which case the generated GSDs would all be balanced. 

\subsection{TFQA vs. NSQA}
\begin{figure*}[htp]
\includegraphics[width=0.99\textwidth]{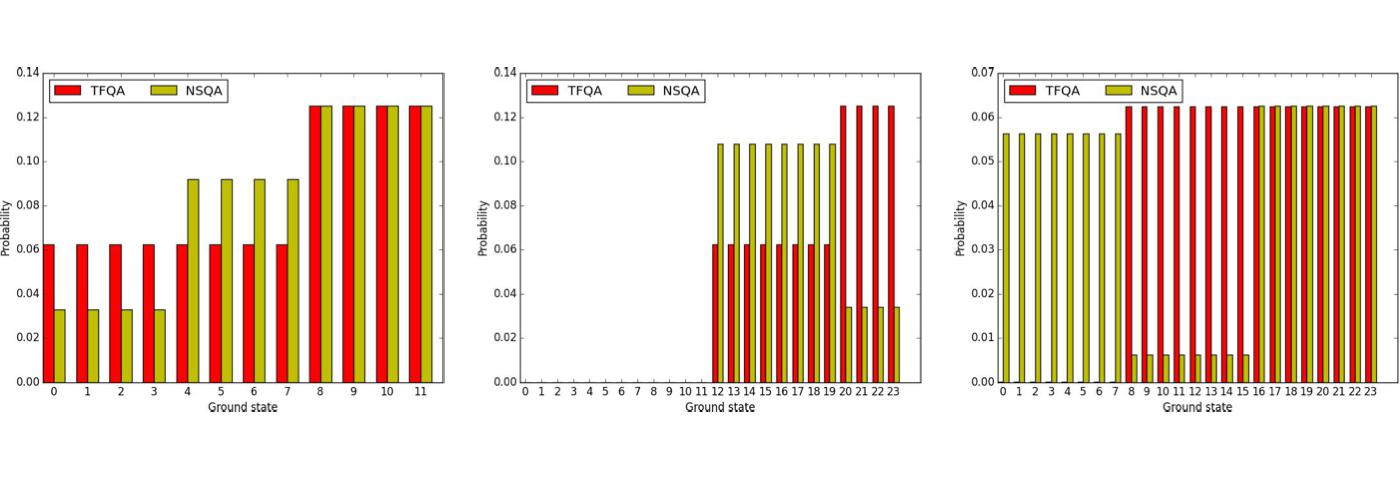}
\caption{{\bf Three sample GSD comparisons between the two drivers: the stoquastic TFQA (red) and non-stoquastic NSQA (yellow).} For the instance on the left both drivers sample the ground-state manifold similarly; for the middle instance less probable configurations for one method become more probable in the other and vice versa; on the right, we find an instance for which states with zero probability of occurring with one type of quantum fluctuations have distinct positive probabilities of occurring in the other.}
\label{fig:drivercomp}
\end{figure*}

Next, we compare the GSDs produced by the ideal zero-temperature quantum annealers, namely the transverse-field annealer (TFQA) against the non-stoquastic annealer (NSQA). As mentioned earlier, for more than half of the instances ($1105$ out of $1909$) the quantum ground-states are found to be `flat' for both annealers (the median bias for both processes was found to be $<10^{-4}$). Discounting for those, we find in general that the chosen driver has a decisive effect on the GSDs. The GSDs of representative instances are shown in Fig.~\ref{fig:drivercomp}, showcasing the tunability of the probability distributions with respect to the `knob' of quantum fluctuations. For the instance on the left, both drivers sample the ground-state manifold similarly; for the middle and right instances, we observe that the driver has a profound effect on the shape of the distribution. For a fraction of the instances ($<5\%$), ground-states that had zero probability for TFQA had strictly positive probability when NSQA was used. That is, the driver often has an effect not only on the probabilities of various states but also on which states have nonzero amplitudes. 

Comparing the bias associated with use of a transverse-field quantum annealer with and without the aid of a non-stoquastic driver, we find that the combination of annealers is far less biased than the transverse-field annealer alone. On average, as shown in Fig.~\ref{fig:QAbiasPlots}(right), NSQA distributions are slightly more biased than TFQA distributions. Nonetheless, as is indicated in the figure, which shows a scatter plot of the bias $b({\bf p}_{\text{TFQA}})$ against the biases of the combination of TFQA with NSQA, there is  merit in `assisted' non-stoquastic ground-state sampling.

\subsection{Experimental DW2 vs. SA and TFQA}
Comparing the distributions generated by the experimental quantum annealer DW2 against SA and against an ideal transverse field quantum annealer (TFQA), we find, as in the other comparisons, only a weak relationship between the output distributions. As summarized in Fig.~\ref{fig:pvalues}, in almost all instances, the KS test yielded a significant difference between the GSDs (also see Table~\ref{table:properties}). 
Figure~\ref{fig:pdgd} shows the GSDs for three representative instances. On the left panel, we find that DW2 as well as SA and TFQA yields an approximately flat probability distribution over the various ground-states. In the middle panel, we find those ground-states that TFQA predicts will appear more often, and indeed appeared more often in the D-Wave experimental GSD. On the right panel, no apparent relationship is found between the three GSDs.  

\begin{figure}[hbp]
{\includegraphics[width=0.99\columnwidth]{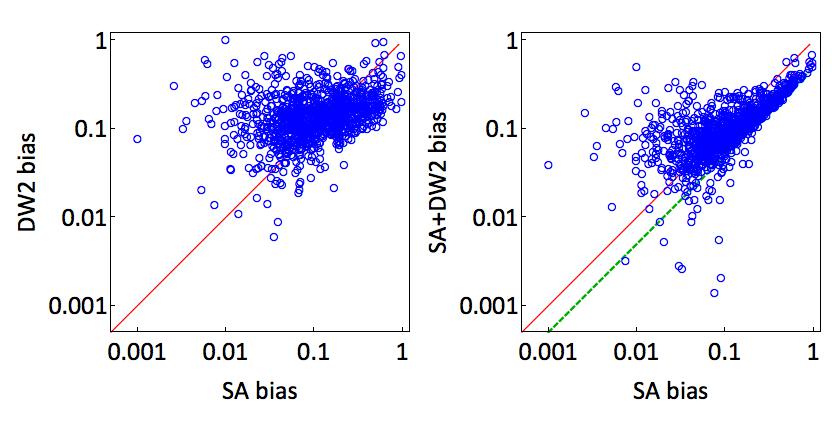}}
\caption{{\bf Scatter plots of SA biases vs. DW2 biases and a combination of SA and DW2.} Left: The scatter plot indicates that the biases of the DW2 processor are similar in nature to those generated by thermal annealing. Right: Merit of using SA in conjunction with an experimental quantum annealer. Here, results are rather mixed. 
While for a large portion (over 50\%) of the instances a combination of the methods leads to a smaller overall bias, i.e., a lesser degree of unfairness, for the rest of the instances the bias is larger. The solid red line is the `equal bias' $y=x$ line, whereas the dashed green $y=x/2$ line is the bias obtained via a combination of the horizontal GSD with a flat unbiased GSD.}
\label{fig:SADW2}
\end{figure}

To understand whether there is merit in using a DW2 processor alongside an SA algorithm for the purpose of producing a more balanced distribution of ground-states, we compare the bias of using SA alone vs. using SA together with DW2. The results are summarized in Fig.~\ref{fig:SADW2}. As the left panel shows, SA and DW2 produce biases with similar distributions. Interestingly, we find that while for some instances use of both methods does reduce the bias, for many others it does not (right panel). When assisted with the DW2 experimental annealer, the median SA bias of $\langle b_{\text{SA}} \rangle =0.10(1)$ remains at $\langle b_{\text{SA+DW2}} \rangle =0.10(1)$.
As we shall discuss next in more detail, while significant differences in the GSDs seem to bode well for the utility of DW2 to generate possible classically suppressed minimizing configurations with high probabilities, the origin of these differences is unclear. 

\section{Discussion}
In this work we studied the capabilities of quantum annealers to
sample the ground-state manifolds of degenerate spin glass
optimization problems. We addressed the question of how differently
ideal zero-temperature and experimental quantum annealers sample the
minimizing configurations of optimization problems than the standard
algorithms, specifically thermal annealers.  Examining both stoquastic
and non-stoquastic quantum fluctuations, we illustrated that
quantum annealers produce, in general,  qualitatively very different
probability distributions than classical annealers; furthermore,
that the final distribution depends heavily on the nature of the
quantum fluctuations. Moreover, we have shown that using quantum
annealers alongside thermal algorithms produces, in general, flatter
distributions of ground-states; that is, a combined use is
significantly more helpful in generating more ground-states than when
using classical algorithms alone.

An earlier work by Matsuda et al.~\cite{matsuda} is worth mentioning here, where a five qubit example has been studied to  show that usage of a transverse-field QA may result in the uneven sampling of a classical ground state manifold. Specifically, the authors found that some classical ground state configurations were unreachable via that driver, whereas more sophisticated drivers that ensured that any two states in the computational basis had non-vanishing matrix elements, resolved that matter. A major caveat pointed out by the authors of Ref.~\cite{matsuda}, though, was that the non-locality of their enhanced driver Hamiltonian had made it very difficult to study numerically at large system sizes (in addition, non-local terms are also known to be problematic to implement physically). We also note an experimental study of the newer-generation DW2X chip~\cite{exponentiallyBiased} where sampling biases have been observed as well. Our study indeed demonstrates that modifying the driver may be useful. In contrast to the above studies, we have studied non-stoquastic yet local (i.e., two-body) driver Hamiltonians, that as such are physically more reasonable. While we find that non-stoquastic local drivers do indeed produce different GSDs than transverse-field drivers, the manner in which they sample the ground state manifolds is not found to be particularly more uniform than with the standard driver.

While we have not explicitly discussed in this work the \emph{speed}
in which these distributions are obtained, we have demonstrated, for
the first time to our knowledge, that annealers driven by quantum
fluctuations may be used to assist existing traditional algorithms in
finding all, or as many as possible, minimizing configurations of
optimization problems.  That quantum annealers may obtain
`classically rare' minimizing configurations has numerous immediate
applications in various fields, such as k-SAT filtering, detecting
hardware faults and verification and validation (V\&V).

An interesting question that arises at this stage and which we believe warrants further investigation is how one should choose the strength or
nature of the quantum fluctuations to boost the probabilities of
classically suppressed configurations. Presumably, algorithms that
adaptively control the nature of the driver terms based on repeated
applications of the annealing process would be an interesting path to
explore (see, e.g., Ref.~\cite{dickson:12}).
Another question that remains open has to do with the nature of the GSDs
generated by the DW2 experimental quantum annealer. As we have shown,
for most instances, the experimentally generated distributions
are only weakly correlated with those of the classical thermal ones,
nor have these been found to correlate with the distributions obtained
by the zero-temperature ideal transverse-field quantum annealers.  The dominant
mechanisms that determine the distributions emerging from the D-Wave
devices remain unknown, however.  Specifically, the question of whether
the yielded distributions are `quantum' in nature---i.e., dominated by
the quantum fluctuations, or classical, i.e., set by thermal
fluctuations, or simply by intrinsic control errors (ICE)---is still
left unanswered. One plausible explanation is that the generated
distributions are a result of a combination of thermal and quantum
fluctuations, given the finite-temperature of the device.  A recent
conjecture\cite{QBM} suggests that these experimental devices in fact
sample from the quantum ground-state at a point midway through the
anneal at a so-called `freeze-out' point (or region) where thermal
fluctuations become too weak to generate any changes to the
state. Another plausible scenario is that the generated
distributions are determined by intrinsic control errors---those
errors that stem from the analog nature of the processor and as such
may have a decisive effect on the resulting ground-state
distributions\cite{martinMayor:15}.  If the latter conjecture is
found to be true, then it could be that these differences in GSDs are easily
reproducible by classical means by simply injecting random errors to
the problem parameters.

Finally, this study raises an interesting question concerning the computational complexity of faithfully sampling the quantum ground-states of spin glass Hamiltonians. While for (ideal) quantum annealers the successful sampling of the quantum ground-state consists solely of performing an adiabatic anneal followed by a computational basis measurement, classical algorithms must follow a different path to accomplish the same task. 
Since non-stoquastic Hamiltonians cannot be efficiently simulated, the prevailing algorithm to date for sampling the quantum ground-state of such Hamiltonians is one which uses degenerate perturbation theory. 
The latter technique consists of first diagonalizing the problem Hamiltonian to find all its (classical) ground-states, followed by an application of degenerate perturbation theory (up to $N$-th order in the worst case). This last step consists of tracking, in the worst case, all $N$-spin paths from any one classical ground-state to any other. At least naively,  the computational complexity involved in doing so scales as $N!$. Thus, the problem of sampling from quantum ground-states generated by non-stoquastic driver Hamiltonians may eventually prove to be a path to quantum enhancements of a novel kind.

\section*{Methods}

To test the capabilities of quantum annealers tasked with the identification of the individual ground-state configurations
of a given problem versus classical algorithms, we first generate random spin glass instances for which the enumeration of the minimizing configurations is feasible. This a priori requirement allows for the proper evaluation of the sampling capabilities of the tested algorithms in an unbiased and consistent manner. 

\subsection{Generation of Instances}
For the generation of instances, in this work we have chosen to study problems constructed around `planted solutions'---an idea borrowed from constraint satisfaction (SAT) problems\cite{Barthel:2002tw,Krzakala:2009qo}. In these problems, the planted solution represents a ground-state configuration of Eq.~\eqref{eq:Hp} that minimizes the energy and is known in advance. The Hamiltonian of a planted-solution spin glass is a sum of terms, each of which consists of a small number of connected spins, namely, $H=\sum_j H_j$\cite{hen:15}. Each term $H_j$ is chosen such that one of its ground-states is the planted solution. It follows then that the planted solution is also a ground-state of the total Hamiltonian, and its energy is the ground-state energy of the Hamiltonian. 
Knowing the ground-state energy in advance circumvents the need to verify the ground-state energy using exact (provable) solvers, which rapidly become too expensive computationally as the number of variables grows. The interested reader will find a more detailed discussion of planted Ising problems in Ref.\cite{hen:15}. As we show next, studying this type of problem also allows us to devise a practical algorithm capable of finding all minimizing configurations of the generated instances. 

\subsection{Enumeration of Minimizing Configurations}
\begin{figure}
{\includegraphics[width=0.9\columnwidth]{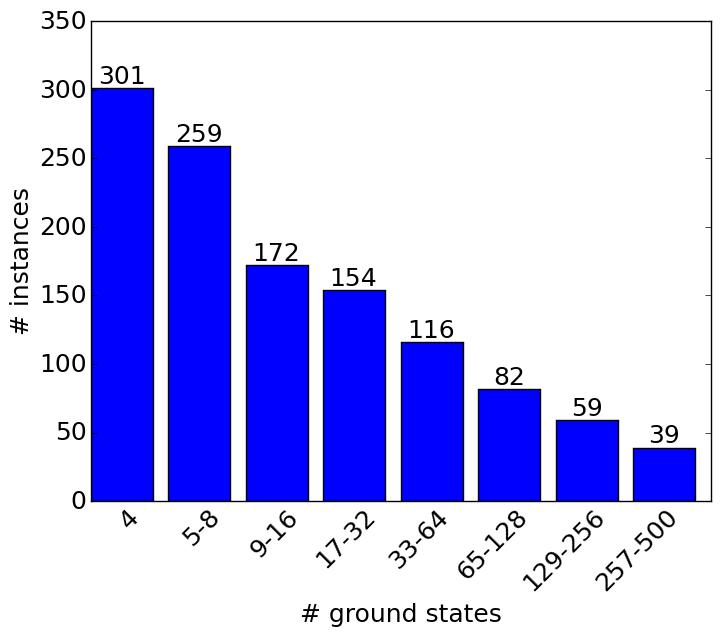}}
\caption{{\bf Histogram of number of problem instances vs. number of ground-states.} The distribution of number of solutions is a function of clause density after discarding instances with more than $500$ solutions.}
\label{fig:nsol}
\end{figure}

The enumeration of all minimizing configurations of a given optimization problem is a difficult task in the general case, belonging to the complexity class $\#P$. The exhaustive search for all solutions of an $N$-spin Ising problem becomes unfeasible for $N>40$ bit problems as the search space grows exponentially with the size of the problem. 
To successfully address this difficulty, we use the fact that our generated instances consist of a sum of terms---each of which has all minimizing configurations as its ground-state (these are commonly referred to as frustration free Hamiltonians). To enumerate all minimizing configurations, we implement a form of the 'bucket' algorithm\cite{dechter} designed to eliminate variables one at a time to perform an exhaustive search efficiently (for a detailed description of the algorithm, see \appcsa). Implementing the bucket algorithm and running it on the randomly generated planted-solution instances, discarding instances with more than $500$ solutions, has yielded the histogram depicted in Fig.~\ref{fig:nsol} which provides the number of problem instances used in this study versus number of ground-states.

\subsection{Calculation of the Quantum Ground-state}
Here we review the algorithm used to compute the `quantum ground-state' of $H_p$, namely, the $s$-dependent ground-state in the limit $s\to 1$ (from below) of the $s$-dependent Hamiltonian,
\beq
H(s)=(1-s) H_d + s H_p  \,.
\eeq
We consider an ideal zero-temperature adiabatic evolution that starts with the driver $H_d$ 
(e.g., a transverse-field Hamiltonian) at $s=0$, and ends with the problem Hamiltonian $H_p$ at $s=1$, where the problem Hamiltonian encodes a classical cost function with $D$ degenerate ground-states. 

Calculating the quantum ground-state in the presence of a degenerate
ground subspace of an $N \approx 500$-qubit problem Hamiltonian is a nontrivial task.  In general, it requires the
diagonalization of $H(s)$ which scales exponentially with the number of spins. 
Our workaround combines the Rayleigh-Ritz variational principle, taking advantage of the fact the sought state is that of minimal energy, combined with degenerate perturbation theory (see e.g., Ref.\cite{galindo:91}), in which $H_d$ is considered a perturbation to $H_p$.

We begin by observing that perturbation theory separates the quantum ground-state from the other states in the limit $s \to 1$ (as shown for example in Fig.~\ref{fig:example}) using a growing sequence of sets of states $S_k$ and subspaces $V_k$ such that: 
\begin{enumerate}
\item The set $S_1=\{|\phi_1\rangle, |\phi_2\rangle,\ldots |\phi_D\rangle\}$  contains all the classical ground-states. 
\item Subspace $V_1$ is the linear span of
  $S_1$ (i.e., the set of all linear combinations of vectors in $S_1$).
\item The states $|\phi\rangle \in S_k$ meet three requirements, (a)
  $|\phi\rangle$ is orthogonal to $V_{k-1}$, (b) $|\phi\rangle$ belongs to
  the computational basis and hence is an eigenstate of $H_p$, and
  (c) $|\phi\rangle$ has a non-vanishing matrix element $\langle \phi|
  H_d^{k-1}|\phi_i\rangle$ for some classical ground-state
  $|\phi_i\rangle\in S_1$.
\item $V_k$ is the linear span of the union
$S_1\cup S_2\cup \ldots \cup S_k$.
\end{enumerate}
For instance, for the transverse-field driver Eq.~\eqref{eq:hdtf}, the set $S_2$ is
composed of the states obtained from each classical ground-state by flipping
only one bit (these new states must not be ground-states themselves).

Additionally, we take advantage of the symmetry of our Hamiltonians $H(s)$
under global bit flip [see
  Eqs.~\eqref{eq:Hp},~\eqref{eq:hdtf} and~\eqref{eq:def-hdnsq}]. For any
vector in the computational basis $|\phi\rangle$, we denote the state obtained by flipping all bits in $|\phi\rangle$ as $|\bar
\phi\rangle$, and
the symmetric ($+$) and antisymmetric ($-$) components of $|\phi\rangle$
as $(|\phi\rangle \pm |\bar \phi\rangle)/\sqrt{2}$. Similarly, since $H_d$
is also symmetric under global bit-flip, the subspaces $V_k$ split into
symmetric and antisymmetric subspaces $V_k=V_k^\text{S}\oplus
V_k^\text{A}$. We shall therefore only consider $V_k^\text{S}$ from now on.

Perturbation theory prescribes in a well-defined manner the specific
linear combination of states in $V_k^\text{S}$ that compose the ground-state to $k$-th order. However, this prescription is
contrived. To simplify the procedure, we  turn to the
Rayleigh-Ritz variational principle. Restricting our test
functions to the subspace $V_k^\text{S}$, the ground-state can also be defined as
\beq\label{eq:RR} 
|\psi_{GS}(s)\rangle =
\arg\min_{\psi\in V_k^\text{S}} \frac{\langle \psi |
  H(s) | \psi \rangle} {\langle \psi|\psi \rangle} \,,
\eeq
i.e., the state with least energy. 
Our variational estimate is consistent with the $k$-th order perturbation
theory because $V_k^\text{S}$ contains all the states 
relevant to $k$-th order perturbation. The minimizing state
$|\psi_{GS}(s)\rangle$ can be easily found using a conjugate gradient
method (for real Hamiltonians, such as ours, the search for
  the minimum in Eq.~(\ref{eq:RR}) can be restricted to real states
  \unexpanded{$|\Psi\rangle$}; our conjugate gradient follows Appendix
  A of Ref.\cite{hen:12}).

Our algorithm proceeds as follows. Considering initially only the order $k=1$ in
Eq.~\eqref{eq:RR}, we set $s=s_*$, a small number such that the ground-
state of $H(s_*)$ is close to the ground-state of $H_d$ and for which
$H_p$ serves as a perturbation (we set $s_*=0.1$) and minimize $H(s_*)$. We then verify that the minimizing state is not degenerate. To do that, it is sufficient to run the conjugate gradient
  minimization routine twice, starting from two independent random states and checking that
  the two obtained ground-states are linearly dependent. Finding the same ground-state twice in the presence of degeneracies
  is a zero-measure event as the minimization is that of a quadratic function, for which there are no local maxima separating different minima. If the minimizing
state is found to be unique, we proceed to slowly increase
$s$ up to $s\to 1$ minimizing the wave-function along the way If on the other hand the degeneracies are not broken at $s=s_*$, we
expand the variational search to $V_2^\text{S}$ and then proceed with
the numerical annealing all the way to $s\to 1$. The state obtained at the end of this procedure is the quantum ground state of $H_p$ as driven by $H_d$.

In the case of $H_d^{\text{TF}}$ we found that out of about
$1900$ randomly generated spin glass instances, considering
$V_1^\text{S}$ sufficed to find the quantum ground-state for $912$ of them, and the inclusion of
$V_2^\text{S}$ sufficed for the rest.  In the non-stoquastic
$H_d^{\text{NS}}$ case, $1508$ were solved with $V_1^\text{S}$, and
the rest with $V_2^\text{S}$. We note that for those instances in which all degeneracies are removed at the level of
$V_1^\text{S}$, the numerical annealing can be dispensed of altogether.
This is because the problem Hamiltonian $H_p$, when restricted to $V_1^\text{S}$, is merely the classical
ground-state energy times the identity.

\subsection{Comparing Empirical Distribution Functions} 
Here we discuss the statistical comparisons of pairs of GSDs over the ground-state manifold of a spin glass with $D$ solutions, where at
least one of the distributions is empirically estimated via
experiment or simulation. Let the two probability distributions be denoted ${\bf p}^{(1)}$, $ {\bf p}^{(2)}$ where $p^{(1)}_i$ and
$p^{(2)}_i$ are the probabilities of obtaining ground-state $i$ with the first and second method, respectively. Let $N_1$ and $N_2$ be the sample sizes
(i.e., number of anneals) for these two experiments (if one of the
methods generates an analytic probability distribution, this
corresponds to taking its number of anneals to infinity). Hence, if
ground-state $i$ was found $n_i^{(1)}$ times with method one, then
$p^{(1)}_i=n_i^{(1)}/N_1$ (with similar definitions for method two). It follows that the square of the $\chi^2$-distance between distributions ${\bf
  p}^{(1)}$ and $ {\bf p}^{(2)}$ is\cite{press:95}
\beq
\label{eq:chi2-def}
\Vert \mathbf{p}^{(1)}-\mathbf{p}^{(2)}  \Vert^2_{\chi^2} = \sum_{i=1}^D 
\frac{\left(\sqrt{\frac{N_1}{N_2}}\,n_i^{(2)}\ -\ \sqrt{\frac{N_2}{N_1}}\,n_i^{(1)}\right)^2}{n_i^{(1)}+n_i^{(2)}}\,.
\eeq
If one of the two probabilities, say $ {\bf p}^{(2)}$, is known analytically,
we may just take the limit $N_2\to\infty$ in Eq.~\eqref{eq:chi2-def}, obtaining
\beq
\label{eq:chi2-one sided-def}
\Vert \mathbf{p}^{(1)}-\mathbf{p}^{(2)} \Vert^2_{\chi^2, \text{one
    sided}} = \sum_{i=1}^D \frac{ \Big( N_1
  p_i^{(2)}\ -\ n_i^{(1)}\Big)^2}{N_1 p_i^{(2)}}\,.  \eeq The
Kolmogorov-Smirnov test proceeds as follows. Here, for the null
hypothesis, one assumes that both ${\bf p}^{(1)}$ and $ {\bf p}^{(2)}$
are drawn from the same underlying probability distribution. If the
null hypothesis holds, then for large $N_1$, $N_2$ and $D$, the
probability distribution for $\Vert \mathbf{p}^{(1)}-\mathbf{p}^{(2)}
\Vert^2_{\chi^2}$ can be computed assuming Gaussian statistics. Hence,
one would just compare the $\Vert \mathbf{p}^{(1)}-\mathbf{p}^{(2)}
\Vert^2_{\chi^2}$ computed from empirical data with the known
$\chi^2$-distribution. This comparison yields the probability (or
$p$-value) that the null hypothesis holds\cite{press:95}.

In our case, neither $N_1$, $N_2$ nor $D$ are exceedingly
large. Therefore, rather than relying on a precomputed
$\chi^2$-probability, we resort to using a bootstrapped version
of the Kolmogorov-Smirnov (KS) test: assuming for the null hypothesis that the probability distributions
associated with the two methods are in fact identical, the underlying
distribution $p$ is given by: \hbox{$ p_i = (N_1 p^{(1)}_i + N_2
  p^{(2)}_i)/(N_1 + N_2) $} (this $p_i$ is our best guess for the probability
of getting the $i$-th ground-state given the null hypothesis). As a next step, we simulate
synthetic experiments based on this underlying distribution. Each
synthetic experiment consists of extracting $N_1$ ground-states to
form a synthetic probability distribution ${\bf P}_1$, and $N_2$ ground-states to form a probability distribution ${\bf P}_2$ (if a method is
analytic then we set ${\bf P}_i= {\bf p}_i$ by law of large numbers). We
simulate a large number of such synthetic experiments, and measure the
proportion of experiments for which we find $\norm{{\bf P}_1-{\bf P}_2} >
\norm{{\bf p}^{(1)} - {\bf p}^{(2)}}$. This proportion corresponds to the
$p$-value for our test.
As a check, we have compared our $p$-values with the ones obtained
using the tabulated $\chi^2$ probability verifying that the
$p$-values of the two tests agree to within a few percent.

\section*{Acknowledgements}
This research used resources of the Oak Ridge Leadership Computing Facility at the Oak Ridge National Laboratory, which is supported by the Office of Science of the U.S. Department of Energy under Contract No. DE-AC05-00OR22725.
Computation for the work described here was also supported by the University of Southern California's Center for High-Performance Computing (\url{http://hpcc.usc.edu}).

\bibliography{refs}


\pagebreak

\section{Supplementary Information}

\subsection{\label{app:dwave} The D-Wave Two Quantum Annealer}

\begin{figure}[hbp]
\begin{center}
\includegraphics[width=0.7\columnwidth]{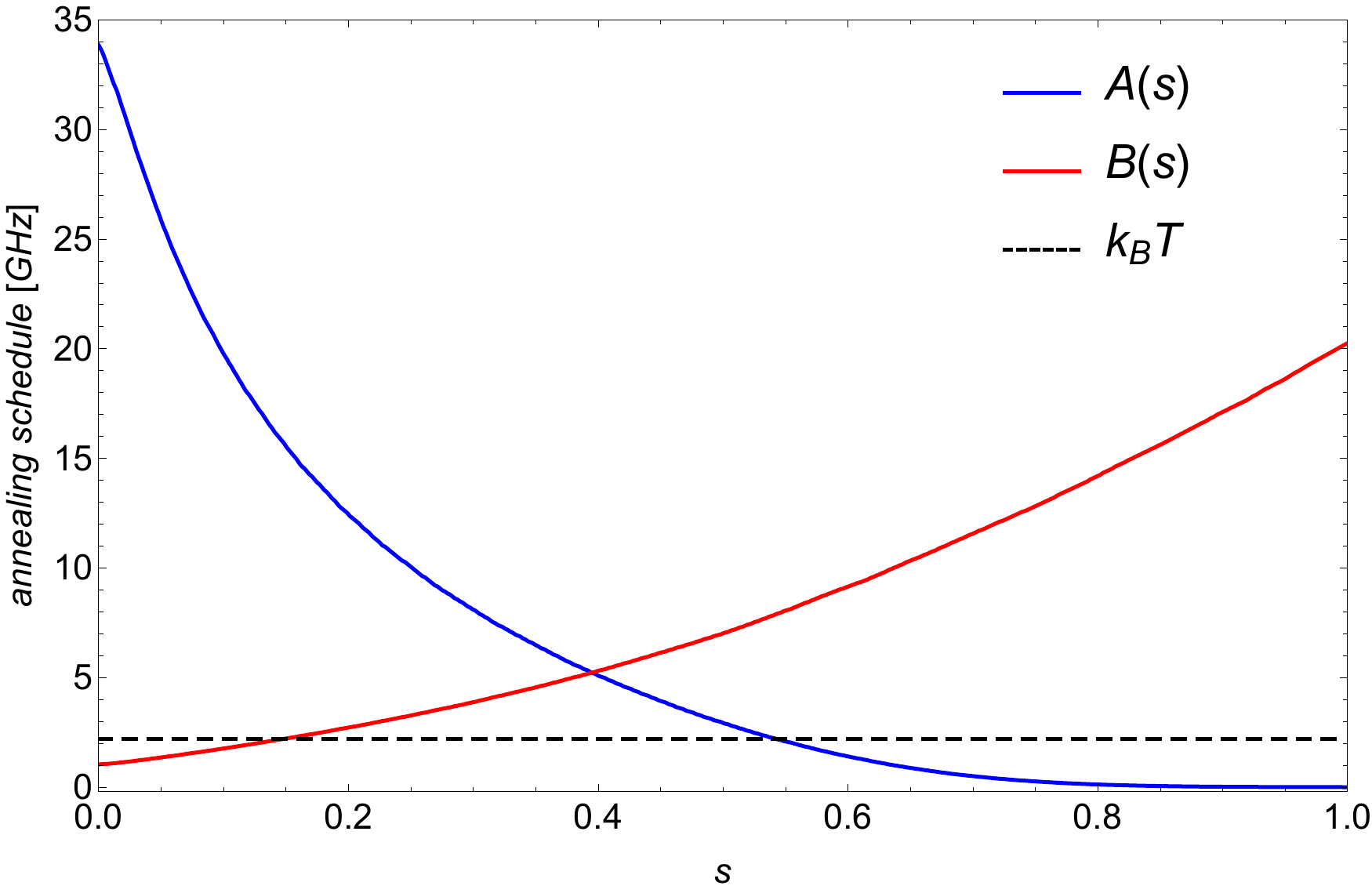}
\caption{\textbf{Annealing schedule of the DW2.}  The annealing curves $A(s)$ and $B(s)$ with $s=t/\mathcal{T}$ are calculated using rf-SQUID models with independently calibrated qubit parameters. Units of $\hbar = 1$.  The operating temperature of $17$mK is also shown as a dashed line.}\label{fig:schedule}
\end{center}
\end{figure}

\begin{figure}[htp]
\begin{center}
\includegraphics[angle=270,trim={3.75cm 3cm 3cm 3cm},clip,width=0.7\columnwidth]{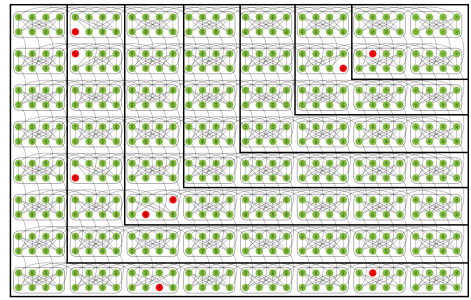}
\caption{\textbf{The DW2 Chimera graph.} The qubits or spin variables occupy the vertices (circles) and the couplings $J_{ij}$ are along the edges. 
Of the $512$ qubits, $504$ were operative in our experiments (green) and $8$ were not (red). 
}
\label{fig:chimera}
\end{center}
\end{figure}

The D-Wave Two (DW2) processor is marketed by D-Wave Systems Inc. as a quantum annealer. 
The annealer evolves a physical system of superconducting flux qubits according to the time-dependent Hamiltonian
\beq \label{eq:Hquantum}
H(t) = A(t) \sum_{i} H_d^{\text{TF}} + B(t) H_p, \quad t\in[0,\mathcal{T}] \,.
\eeq
The problem Hamiltonian $H_p$ is given by
\beq
H_{p}=\sum_{\langle ij\rangle} J_{ij} \sigma^z_i \sigma^z_j + \sum_i h_i \sigma^z_i \,, 
\eeq 
where $\sigma_i^z$ is the Pauli Z spin-$1/2$ matrix acting on spin $i$, the set $\{J_{ij}\}$ are programmable parameters
and  $\langle ij\rangle$
sums over the edges of an $N=504$-qubit Chimera graph---the hardware graph of the processor.
The transverse field driver Hamiltonian $H_d^{\text{TF}}$ is given by
\beq
H_d^{\text{TF}} =- \sum_i \sigma_i^x \,,
\eeq
where $\sigma_i^x$ is the Pauli X spin-$1/2$ matrix acting on spin $i$. 
The annealing schedules given by $A(s)$ and $B(s)$ are shown in Fig.~\ref{fig:schedule} as a function of the dimensionless parameter $s=t/\mathcal{T}$.  Our experiments used the DW2 device housed at University of Southern California's Information Sciences Institute, which is held at an operating temperature of $17$mK. 
The Chimera graph of the DW2 used in our work is shown in Fig.~\ref{fig:chimera}. Each unit cell is a balanced $K_{4,4}$ bipartite graph. In the ideal Chimera graph (of $512$ qubits) the degree of each vertex is $6$ (except for the corner cells). In the actual DW2 device, only $504$ qubits are functional. 

\subsection{\label{app:csa}Constraint Solver Algorithm}

This is a description of the algorithm used to search for and list all solutions, or minimizing configurations, of a spin glass instance with a planted solution. The algorithm takes as input a specified set of constraints, namely, sets of bit assignments, each of which minimizes the local Ising Hamiltonians' $H_j$, where the total planted-solution Hamiltonian is $H=\sum_j H_j$. Each such term is minimized by a finite set of spin configurations (involving only those spins on which the local Hamiltonian is defined). 
The algorithm is an exhaustive search and is an implementation of the bucket elimination algorithm described in Ref.~\cite{dechter}.

The problem structure consists of a set of Ising spins (equivalently, bits) where each value may be $+1$ or $-1$.  
A set of constraints restricts subsets of the bits to certain values. 
The goal is to assign values to all the bits so as to satisfy all the constraints.
Each constraint (in this case, the optimizing configurations of the $H_j$ Hamiltonians) applies to a subset of the bits. 
Each constraint contains a list of allowed settings for that subset of bits. 
The constraint is met if the values of the bits in the subset matches one of the allowed settings.
The constraints for a particular problem are read from a text file. A sample input for a single constraint is:
\begin{small}
\begin{center}
\begin{tabular}{|c || r | r| r |r| r| r| r| r| r| r| r| r||}
\hline
bit & sol. & sol.  & sol. & sol. & sol. &  sol. &  sol. &sol. &  sol. &  sol.  &  sol.   &sol. \\
 index & \#1 &  \#2  & \#3 &  \#4 &  \#5 &  \#6 &  \#7  & \#8&  \#9 &  \#10 &   \#11  & \#12\\
\hline
480   &  1      & -1&      1    &   -1 &     1       & -1&      1     &  -1 &     1     &  -1 &     1    &   -1 \\
485    & -1     & 1   &    1   &    -1   &   1      & -1  &    1     &  -1    &  1     &  -1   &   1    &   -1 \\
493    & 1      & -1    &  1 &      -1    &  -1    &  1   &    -1    &  1     &  -1    &  1    &   -1   &   1 \\
491     &-1    &  1       &-1   &   1      & -1   &   1   &    1    &   -1    &  1      & -1   &   1    &   -1 \\
495     &1   &    -1  &    1  &     -1     & 1   &    -1   &   1    &   -1     & -1    &  1     &  -1   &   1 \\
487    & 1 &      -1    &  1&       -1     & 1 &       -1  &    1 &       -1    &  1 &      -1    &  -1 &     1 \\
\hline
\end{tabular}
\end{center}
\end{small}
The first column is the bit numbers, six in total. 
The following columns are allowed values for this subset of bits indicating that the term has 12 minimizing bit assignments. 
This input specifies $12$ allowed settings for bits $480, 485, 493, 491, 495$ and $487$.

The bucket elimination algorithm, as applied to this problem, consists of the following steps:
\begin{verbatim}
First, determine if solutions exist:
  while(true)
    select a bit to eliminate
      if no more bits:
      		 exit loop
      save constraints which contain selected bit
      combine all constraints containing bit
      generate new constraints without bit
      if combined constraints have contradiction:
      		 exit loop
  end while
\end{verbatim}

This algorithm is guaranteed to find all solutions, but may exceed time and memory limitations. 
All steps are well-defined except for the step which selects the next bit to eliminate. 
The order in which bits are eliminated dramatically affects the time and memory required. 
Determination of the optimal order to eliminate bits is known to be NP complete.


Assume there exists a list of constraints, as described above, each of which contains a specified bit. 
 For simplicity, first assume there are only two such constraints. 
 For example, consider the following two simple constraints, both of which contain bit $\#1$:
\begin{center}
 \begin{tabular}{||c |c || r |r| r||}
 \hline
constraint \#1:     &      bits         &\#1 &  \#2 &   \#3 \\
\hline
      allowed settings: &  $(1_a)$ & -1  & -1 &  -1 \\
                                    &    $(1_b)$ & +1  & +1 & -1     \\     
\hline
constraint \#2:          & bits   &         \#1&   \#2 &  \#4 \\
\hline
      allowed settings:   & $(2_a)$  & -1 &  -1 &  -1 \\
                                      &     $(2_b)$ & +1 &  -1 &  -1 \\
 \hline
\end{tabular}
 \end{center}
 
To combine the two constraints, combine each of the allowed settings in the first constraint with each of the allowed settings in the second constraint. 
 When combining two settings, any bit that is in both settings must agree, or no new constraint is generated. 
 The bit chosen for elimination is not contained in the newly generated constraint. 
 If at the end, no new constraints have been generated, a contradiction exists which prevents these constraints from mutual satisfaction. 
 This indicates that the original problem has no solutions.

Application of this step to constraints $1_a, 1_b, 2_a$ and $2_b$ above requires four steps. 
 The bits contained in the newly created constraints contain the union of the bits in the constraints, minus bit $\#1$, the bit being eliminated.
\begin{center}
 \begin{tabular}{|c|c|}
 \hline
                                   &     combined configuration\\
 \hline
     combine $1_a$ and $2_a$    &   $(-1,   -1,   -1)$ \\
     combine $1_a$ and $2_b$     &  empty -- bit $\#1$ differs in the two \\
     combine $1_b$ and $2_a$    &   empty -- bit $\#1$ differs in the two \\
     combine $1_b$ and $2_b$     &  empty -- bit $\#2$ differs in the two \\
 \hline
\end{tabular}
 \end{center}
 
The new constraint consists of a single entry for bits $\#2$, $\#3$, and $\#4$ set to $(-1, -1, -1)$. 
 To combine more than two constraints, the first two are combined, then the result is combined with the next constraint and repeated until all constraints have been combined. The process of combining constraints can cause the number of allowed bit set values to shrink, as in the example above, or to grow.

The amount of time and memory required to eliminate all bits from the original set of constraints is highly dependent on the order in which bits are eliminated from the original set of constraints.  Many heuristics were tested to select the best next bit to be eliminated.
No deterministic algorithm was found that yielded acceptable time and memory use on all of the input data sets. The approach that was ultimately found to be effective was to use a combination of heuristics and randomness to select the next bit. Each time an elimination bit is to be chosen, one of six heuristics is chosen at random. The six heuristics are different functions of the number of unique bits in the constraints to be combined, the maximum number of solution sets in any of the constraints to be combined, and the sum of the number of solution sets in the constraints to be combined.

Solutions are enumerated by iterating over the saved variable tables in reverse order. The last table contains the allowed values for the last variable eliminated. Each of these values can be substituted into the previous table to generate allowed value sets for the last two variables. This process is repeated for each table until allowed value sets for all variables are generated. The list is truncated at each step if the number of value sets exceeds a specified limit.

\end{document}